\documentclass[11pt]{article}
\usepackage{authblk}
\usepackage[]{inputenc}
\usepackage{amsfonts}
\usepackage{latexsym,amsmath}
\usepackage{amssymb,array}
\makeatletter \oddsidemargin 0in \evensidemargin 0in \textwidth 16cm
\RequirePackage[dvips]{graphicx} \textheight 20cm
\begin{document}
\newcommand{\om}{\omega}
\newtheorem{thm}{Theorem}[section]
\newtheorem{pro}{Proposition}[section]
\newtheorem{rem}{Remark}[section]
\newtheorem{ce}{Counterexample}[section]
\newtheorem{cor}{Corollary}[section]
\newtheorem{d1}{Definition}[section]
\newtheorem{ex}{Example}[section]
\newtheorem{lem}{Lemma}[section]
\numberwithin{equation}{section}
\title{On Relative Ageing of Coherent Systems with Dependent Identically Distributed Components}
\author[a]{Nil Kamal Hazra\footnote{Corresponding author, email: nkhazra@iiitdm.ac.in}}
\author[b]{ Neeraj Misra}
\affil[a]{Department of Mathematics, Indian Institute of Technology Jodhpur, Karwar $342037$, India}
\affil[b]{Department of Mathematics and Statistics, Indian Institute of Technology Kanpur, Kanpur 208016, India}
\date{}
\maketitle
\begin{abstract}
Relative ageing describes how a system ages with respect to another one. The ageing faster orders are the ones which compare the relative ageings of two systems. Here, we study ageing faster orders in the hazard and the reversed hazard rates. We provide some sufficient conditions for proving that one coherent system dominates another system with respect to ageing faster orders. Further, we investigate whether the active redundancy at the component level is more effective than that at the system level with respect to ageing faster orders, for a coherent system. Furthermore, a used coherent system and a coherent system made out of used components are compared with respect to ageing faster orders.
\end{abstract}
{\bf Keywords:} Coherent system, dual distortion/domination function, $k$-out-of-$n$ system, redundancy, residual lifetime, stochastic orders
\\{\bf 2010 Mathematics Subject Classification:} Primary 90B25
\\\hspace*{3.2 in}Secondary 60E15; 60K10
\section{Introduction and Preliminaries}\label{se1}
Ageing is a common phenomenon experienced by both living organisms and mechanical systems. It largely describes how a system/living organism improves or deteriorates with age. The study of stochastic ageing has receieved considerable attention from researchers in the last few decades. In the literature, many different types of stochastic ageing concepts (e.g., increasing failure rate (IFR), increasing failure rate in average (IFRA), etc.) have been developed to describe different ageing characteristics of a system. There are three types of ageings, namely, positive ageing, negative ageing and no ageing. A brief discussion on different ageing concepts could be found in Barlow and Proschan~\cite{bp}, and Lai and Xie~\cite{lx0}. Similar to these ageing concepts, there is another useful notion of ageing, called relative ageing, which describes how a system ages relative to another one.
\\\hspace*{0.2 in}The proportional hazard (PH) rate model, commonly known as Cox's PH model (see Cox~\cite{cox}), is widely used to analyze the failure time data in reliability and survival analysis. Later, different other models were introduced, namely, proportional mean residual lifetime model, proportional reversed hazard rate model, proportional odds model, etc (see Marshall and Olkin~\cite{mo}, Lai and Xie~\cite{lx0}, and Finkelstein~\cite{f8}). In many real life scenarios, the phenomenon of crossing hazards/mean residual lives has been observed (see Pocock et al.~\cite{pgk}, Champlin et al.~\cite{cmeg}, and Mantel and Stablein~\cite{ms}). To handle this crossing hazard rates problem, Kalashnikov and Rachev~\cite{kr} introduced a stochastic order (called ageing faster order in the hazard rate) based on the concept of relative ageing. Indeed, this approach could be considered as a reasonable alternative to the PH model. A detailed study of this order is done by Sengupta and Deshpande~\cite{sd}. In addition, they have also introduced two other similar kinds of stochastic orders. Later, Finkelstein~\cite{f6} proposed a stochastic order, based on mean residual lifetime functions, that describes the relative ageings of two life distributions, whereas Rezaei~\cite{rgi} introduced a similar stochastic order in terms of the reversed hazard rate functions. Some generalized orderings in this direction were proposed by Hazra and Nanda~\cite{hn0}.
\\\hspace*{0.2 in}The basic structures of most of the real life systems match with the so called coherent system. A system is called coherent if its all components are relevant and its structure function (see Barlow and Proschan~\cite{bp} for the definition) is monotonically non-decreasing with respect to each argument (which means that an improvement in performance of a component cannot decrease the lifetime of the system). The well known $k$-out-of-$n$ system is a special case of coherent systems. A system of $n$ components is said to be $k$-out-of-$n$ system if it functions as long as at least $k$ of its $n$ components function. Two extreme cases of a $k$-out-of-$n$ system are $1$-out-of-$n$ system (called parallel system) and $n$-out-of-$n$ system (called series system). Further, there is an one-to-one correspondence between a $k$-out-of-$n$ system and an $(n-k+1)$-th order statistic (of lifetimes of $n$ components). Thus, the study of a $k$-out-of-$n$ system is essentially the same as the study of an order statistic. 
\\\hspace*{0.2 in}Stochastic comparisons of coherent systems is considered as one of the important problems in reliability theory. The list of results, so far developed, on various stochastic comparisons of $k$-out-of-$n$ systems with independent components could be found in Pledger and Proschan~\cite{pp}, Proschan and Sethuraman~\cite{ps}, Balakrishnan and Zhao~\cite{bz}, Hazra et al.~\cite{hkfn}, and the references therein. Further, stochastic comparisons of general coherent systems were considered in Esary and Proschan~\cite{ep}, Nanda et al.~\cite{njs}, Kochar et al.~\cite{kms}, Belzunce et al.~\cite{bfrr}, Navarro and Rubio~\cite{nr}, Navarro et al.~\cite{nass1, nass3, npd}, Samaniego and Navarro~\cite{sn}, Amini-Seresht et al.~\cite{azb}, to name a few. Note that all these results are developed using different stochastic orders, namely, usual stochastic order, hazard rate order, likelihood ratio order, etc. However, the study of coherent systems using ageing faster orders are not substantially done yet. Misra and Francis~\cite{mf}, Li and Lu~\cite{ll6}, and Ding and Zhang~\cite{dz} developed some results for $k$-out-of-$n$ systems using ageing faster orders. Later, Ding et al.~\cite{dfz} have given some sufficient conditions in terms of signature to compare the lifetimes of two coherent systems (with independent components) with respect to ageing faster orders. However, there is no such result where the sufficient conditions are given in terms of reliability functions. Furthermore, the coherent systems with dependent components are also not considered yet. Thus, one of our major goals of this paper is to provide some sufficient conditions (in terms of reliability functions) under which one coherent system dominates another one with respect to ageing faster orders.
\\\hspace*{0.2 in}One of the effective ways to enhance the lifetime of a system is by incorporating spares (or redundant components) into the system. Then the key question is $-$ how to allocate spares into the system so that the system's lifetime will be optimum in some stochastic sense? In Barlow and Proschan~\cite{bp}, it is shown that the allocation of active redundancy at the component level (of a coherent system) is superior to that at the system level with respect to the usual stochastic order. Later, many other researchers have studied this problem in different directions (see Boland and El-Neweihi~\cite{be}, Misra et al.~\cite{mdg}, Nanda and Hazra~\cite{nh}, Hazra and Nanda~\cite{hn8}, Zhao et al.~\cite{zzl}, Da and Ding~\cite{dd}, Zhang et al.~\cite{zad}, and the references therein). However, to the best of our knowledge, this problem using ageing faster orders is not studied yet. Thus, another goal of this paper is to derive some necessary and sufficient conditions under which the lifetime of a coherent system with active redundancy at the component level is larger (smaller) than that at the system level with respect to ageing faster orders.
\\\hspace*{0.2 in}The real life systems are either formed by new components or by used components. Consider two coherent systems, namely, a used coherent system (i.e., a coherent system formed by a set of new components, and then the system has been used for some time $t>0$) and a coherent system of used components (i.e., a coherent system formed by a set of components which have already been used for time $t>0$). It is a fact that a coherent system of new components does not always have larger lifetime than a coherent system made out of used components (see Navarro et al.~\cite{nffa}). Similarly, a used coherent system may or may not perform better than a coherent system of used components. The stochastic comparisons between these two systems have been done in numerous papers, see, for example, Li and Lu~\cite{ll}, Gupta~\cite{g}, Gupta et al.~\cite{gmk}, Hazra and Nanda~\cite{hn}, to name a few. However, to the best our knowledge, the ageing faster orders have not yet been used, as a tool, to compare these two systems. Thus, the study of stochastic comparisons between a used coherent system and a coherent system of used components is another thrust area that is to be focused here. 
\\\hspace*{0.2 in}In what follows, we introduce some notation that will be used throughout the paper. For a random variable $W$ (with absolutely continuous cumulative distribution function), we denote its probability density function (pdf) by $f_W(\cdot)$, the cumulative distribution function (cdf) by $F_W(\cdot)$, the hazard rate function by $r_W(\cdot)$, the reversed hazard rate function by $\tilde r_W(\cdot)$ and
the survival/reliability
 function by $\bar F_W(\cdot)$; $\bar F_W(\cdot)=1-F_W(\cdot)$.
\\\hspace*{0.3 in}Let us consider a coherent system with lifetime $\tau\left(\mbox{\boldmath$X$}\right)$ formed by $n$ components having dependent and identically distributed (d.i.d.) lifetime vector $\mbox{\boldmath$X$}=(X_1,X_2,\dots,X_n)$, where $X_i$'s are identically distributed, say $X_i\stackrel{\text{d}}=X,$ $i=1,2,\dots,n$, for some non-negative random variable $X$; here $\stackrel{\text{d}}=$ means equality in distribution.
Then the joint reliability function of $\mbox{\boldmath$X$}$ is given by
\begin{eqnarray*}
\bar F_{\mathbf{X}}(x_1,x_2,\dots,x_n)&=&P\left(X_1>x_1,X_2>x_2,\dots,X_n>x_n\right)
\\&=&K\left(\bar F_{X}(x_1),\bar F_{X}(x_2),\dots,\bar F_{X}(x_n)\right),
\end{eqnarray*}
where $K(\cdot,\cdot,\dots,\cdot)$ is a survival copula describing the dependency structure among components of the system. Indeed, this representation is well known through Sklar's Theorem (see Nelsen~\cite{n}). In the literature, many different types of survival copulas have been studied in order to describe different dependency structures among components. Some of the widely used copulas are Farlie-Gumbel-Morgenstern (FGM) copula, Archimedean copula with different generators,
Clayton-Oakes (CO) copula, etc. We refer the reader to Nelsen~\cite{n} for a detailed discussion on the copula theory, and its various applications. In what follows, we give a lemma that describes a fundamental bridge between a system and its corresponding components through the domination function.
\begin{lem}[Navarro et al.~\cite{nass1}]\label{II}
 Let $\tau\left(\mbox{\boldmath$X$}\right)$ be the lifetime of a coherent system formed by $n$ d.i.d. components with the lifetime vector $\mbox{\boldmath$X$}=(X_1,X_2,\dots,X_n)$. Then the reliability function of $\tau\left(\mbox{\boldmath$X$}\right)$ can be written as
 $$\bar F_{\tau\left(\mbox{\boldmath$X$}\right)}(x)=h\left(\bar F_{X}(x)\right),$$
 where $h(\cdot):[0,1]\rightarrow [0,1]$, called the domination (or dual distortion) function, depends on the structure function $\phi(\cdot)$ (see Barlow and Proschan~\cite{bp} for definition) and on the survival copula $K$ of $X_1,X_2,\dots,X_n$. Furthermore, $h(\cdot)$ is an increasing continuous function in $[0,1]$ such that $h(0)=0$ and $h(1)=1$. $\hfill\Box$
\end{lem}
\hspace*{0.2 in}Below we give an example (borrowed from Navarro et al.~\cite{nass1}) that illustrates the result given in the above lemma.
\begin{ex}
Let $\tau\left(\mbox{\boldmath$X$}\right)=\min\{X_1,\max\{X_2,X_3\}\}$, where $\mbox{\boldmath$X$}=(X_1,X_2,X_3)$ is described by the FGM Copula (see Nelsen~\cite{n})
$$K(p_1,p_2,p_3)=p_1p_2p_3(1+\theta(1-p_1)(1-p_2)(1-p_3)), \text{ for }p_i\in(0,1),\; i=1,2,3,\text{ and }\theta \in[-1,1].$$
Further, let $X_1,X_2$ and $X_3$ be identically distributed with a random variable $X$. Then the minimal path sets (see Barlow and Proschan~\cite{bp}) of $\tau\left(\mbox{\boldmath$X$}\right)$ are given by $\{1,2\}$ and $\{1,3\}$. Let $X_{\{1,2\}}$, $X_{\{1,3\}}$ and $X_{\{1,2,3\}}$ be the lifetimes of the path sets $\{1,2\}$, $\{1,3\}$ and $\{1,2,3\}$, respectively. Then the reliability function of $\tau\left(\mbox{\boldmath$X$}\right)$ can be written as
\begin{eqnarray*}
\bar F_{\tau\left(\mbox{\boldmath$X$}\right)}(x)&=&P\left(\{X_{\{1,2\}}>x\}\cup\{X_{\{1,3\}}>x\}\right)
\\&=&P\left(X_{\{1,2\}}>x)+P(X_{\{1,3\}}>x\right)-P(X_{\{1,2,3\}}>x)
\\&=&\bar F_{\mathbf{X}}(x,x,0)+\bar F_{\mathbf{X}}(x,0,x)-\bar F_{\mathbf{X}}(x,x,x)
\\&=&K\left(\bar F_{X}(x),\bar F_{X}(x),1\right)+K\left(\bar F_{X}(x),1,\bar F_{X}(x)\right)-K\left(\bar F_{X}(x),\bar F_{X}(x),\bar F_{X}(x)\right)
\\&=&h\left(\bar F_{X}(x)\right),
\end{eqnarray*}
where
\begin{eqnarray*}
h(p)&=&K(p,p,1)+K(p,1,p)-K(p,p,p)
\\&=&2p^2-p^3-\theta p^3(1-p)^3, \text{ for }p\in(0,1)\text{ and }\theta\in[-1,1]. 
\end{eqnarray*}
\end{ex}
\hspace*{0.2 in}Stochastic orders are commonly used to compare two random variables (or two sets of random variables), and have been extensively studied in the literature due to their various applications in different branches of science and engineering. An encyclopedic information on this topic is nicely encapsulated in the book written by Shaked and Shanthikumar~\cite{ss} (also see Belzunce et al.~\cite{bmr}). For the sake of completeness, we give the following definitions of the stochastic orders that are used in our paper.
\begin{d1}\label{de1}
Let $X$ and $Y$ be two absolutely continuous random variables with cumulative distribution functions $F_X(\cdot)$ and $F_Y(\cdot)$, respectively, supported on $[0,\infty)$. Then $X$ is said to be smaller than $Y$ in
\begin{enumerate}
\item [$(a)$] hazard rate (hr) order, denoted as $X\leq_{hr}Y$, if $${\bar F_Y(x)}/{\bar F_X(x)}\text{ is increasing in } x \in [0,\infty);$$
\item [$(b)$] reversed hazard rate (rhr) order, denoted as $X\leq_{rhr}Y$, if $$ {F_Y(x)}/{ F_X(x)}\text{ is increasing in } x \in [0,\infty).$$
\end{enumerate}
\end{d1}
\hspace*{0.2 in}Similar to the above discussed stochastic orders, there are two more sets of stochastic orders which are useful to describe the relative ageings of two systems. The first set of stochastic orders, known as transform orders (namely, convex transform order, quantile mean inactivity time order, star-shaped order, super-additive order, DMRL order, s-IFR order, etc.), describes whether a system is ageing faster than another one in terms of the increasing failure rate, the increasing failure rate on average, the new better than used, etc. A detailed discussion on these orders could be found in Barlow and Proschan~\cite{bp}, Bartoszewicz~\cite{b}, Deshpande and Kochar~\cite{dk}, Kochar and Wiens~\cite{kw}, Arriaza et al.~\cite{ass}, Nanda et al.~\cite{nhga}, and the refernces therein. The second set of stochastic orders, called ageing faster orders, is defined based on monotonocity of ratios of some reliability measures, namely, hazard rate function, reversed hazard rate function, mean residual lifetime function, etc. For motivation and usefulness of these orders, we refer the reader to Kalashnikov and Rachev~\cite{kr}, Sengupta and Deshpande~\cite{sd}, Di Crescenzo~\cite{d}, Finkelstein~\cite{f6}, Razaei et al.~\cite{rgi}, Hazra and Nanda~\cite{hn0}, Misra et al.~\cite{mfn}, Kayid et al.~\cite{kiz}, and Misra and Francis~\cite{mf2}. Below we give the definitions of the ageing faster orders that are used in our paper.
\begin{d1}
Let $X$ and $Y$ be two absolutely continuous random variables with failure rate functions $r_X(\cdot)$ and $r_Y(\cdot)$, respectively, and reversed failure rate functions $\tilde r_X(\cdot)$ and $\tilde r_Y(\cdot)$, respectively.
Then $X$ is said to be ageing faster than $Y$ in 
\begin{enumerate}
\item [$(a)$] failure rate, denoted as $X\underset{ c}{\prec} Y$, if
$$r_X(x)/r_Y(x)\text{ is increasing in }x\in[0,\infty);$$
\item [$(b)$] reversed failure rate, denoted as $X\underset{ b}{\prec} Y$, if
$$\tilde r_X(x)/\tilde r_Y(x)\text{ is decreasing in }x\in[0,\infty).$$
\end{enumerate}
\end{d1}
 \hspace*{0.2 in}The theory of totally positive functions has various applications in different areas of probability and statistics (see Karlin~\cite{k}). Below we give the definitions of TP$_2$ and RR$_2$ functions. Different properties of these functions are used in proving the main results of our paper.
\begin{d1}
 Let $\mathcal{X}$ and $\mathcal{Y}$ be two linearly ordered sets. Then, a real-valued function $\kappa(\cdot,\cdot)$ defined on $\mathcal{X}\times\mathcal{Y}$,
 is said to be TP$_2$ (resp. RR$_2$) if
 $$\kappa(x_1,y_1)\kappa(x_2,y_2)\geq(\text{resp. }\leq)\;\kappa(x_1,y_2)\kappa(x_2,y_1),$$
  for all $x_1<x_2$ and $y_1<y_2$.$\hfill \Box$
\end{d1}
\hspace*{0.2 in}Throughout the paper increasing and decreasing, as usual, mean non-decreasing and non-increasing, respectively. Similarly, positive and negative mean non-negative and non-positive, respectively. Assume that all random variables considered in this paper are absolutely continuous and non-negative (i.e., distributional support is $[0,\infty)$). By $a\stackrel{\text{sgn}}=b$, we mean that $a$ and $b$ have the same sign, whereas $a\stackrel{\text{def.}}=b$ means that $a$  is defined as $b$. Further, we use bold symbols to represent vectors, and the symbol $\mathbb{N}$ is used to represent the set of natural numbers. We write $\tau_{k|n}$ and $\tau_{l|m}$ to represent the lifetimes of a $k$-out-of-$n$ and a $l$-out-of-$m$ systems, respectively. We use the acronyms $i.i.d.$ and $d.i.d.$ for `independent and identically distributed' and `dependent and identically distributed', respectively.
\\\hspace*{0.2 in}The rest of the paper is organized as follows. In Section~\ref{se2}, we discuss some useful lemmas which are intensively used in the proofs of the main results. In Section~\ref{se3}, we provide some sufficient conditions under which the lifetime of one coherent system is larger than that of an another system with respect to ageing faster orders in terms of the hazard and the reversed hazard rates. In Section~\ref{se4}, we discuss a redundancy allocation problem in a coherent system. We derive some necessary and sufficient conditions under which the allocation of active redundancy at the component level (of a coherent system) is superior to that at the system level with respect to ageing faster orders. Stochastic comparisons between a used coherent system and a coherent system made by used components are discussed in Section~\ref{se5}. The concluding remarks are given in Section~\ref{se6}.
\\\hspace*{0.2 in}All proofs of lemmas and theorems, wherever given, are deferred to the Appendix. 
\section{Useful Lemmas}\label{se2}
In this section we discuss some lemmas which will be used in proving the main results of this paper. In the first lemma we discuss the sign change property of the integral of a function. The following lemma is adopted from Karlin~(\cite{k}, Theorem 11.2, pp. 324-325), and Hazra and Nanda~(\cite{hnx}, Lemma 3.5).
\begin{lem}\label{l1}
 Let $\kappa(x,y)>0$, defined on $\mathcal{X}\times \mathcal{Y}$, be RR$_2$ (resp. TP$_2$), where $\mathcal{X}$ and $\mathcal{Y}$ are subsets of the real line. 
 Assume that a function $f(\cdot,\cdot)$ defined on $\mathcal{X}\times\mathcal{Y}$ is such that
 \begin{itemize}
  \item [$(i)$] for each $x\in \mathcal{X}$, $f(x,y)$ changes sign at most once, and if the change of sign does occur, it is from positive to negative, as $y$ traverses $\mathcal{Y}$;
  \item [$(ii)$] for each $y \in \mathcal{Y}$, $f(x,y)$ is increasing (resp. decreasing) in $x\in \mathcal{X}$;
  \item [$(iii)$] $\omega(x)=\int\limits_{\mathcal{Y}}\kappa(x,y)f(x,y)d\mu(y)$ exists absolutely and defines a continuous function of $x$, where $\mu$ is a sigma-finite measure. 
 \end{itemize}
Then $\omega(x)$ changes sign at most once, and if the change of sign does occur, it is from negative (resp. positive) to positive (resp. negative), as $x$ traverses $\mathcal{X}$.$\hfill\Box$
\end{lem}
\hspace*{0.2 in}In the following lemma we state an equivalent condition of a monotonic function. The proof is straightforward, and hence omitted.
\begin{lem}\label{l2}
 Let $f(\cdot)$ and $g(\cdot)$ be two non-negative real-valued functions defined on $(a,b)\subseteq (0,\infty)$. Then $f(x)/g(x)$ is increasing (resp. decreasing) in $x$, if and only if for any real number $c$, the difference $f(x)-c g(x)$ changes sign at most once, and if the change of sign does occur, it is from negative (resp. positive) to positive (resp. negative), 
 as $x$ traverses from $a$ to $b$.$\hfill\Box$
\end{lem}
\hspace*{0.2 in}Some properties of the reliability functions of a $k$-out-of-$n$ and a $l$-out-of-$m$ systems are discussed in the next two lemmas. Lemma~\ref{l3} ($i$) is obtained in Esary and Proschan~\cite{ep}, whereas Lemma~\ref{l4} ($i$) is obtained in Nanda et al.~\cite{njs}. The other proofs are deferred to the Appendix.
\begin{lem}\label{l3}
Let $h_{k|n}(\cdot)$ and $h_{l|m}(\cdot)$ be the reliability functions of the $k$-out-of-$n$ and the $l$-out-of-$m$ systems with i.i.d. components, respectively, where $1\leq k\leq n$ and $1\leq l\leq m$. Further, let $H_{k|n}(p)=ph_{k|n}'(p)/h_{k|n}(p)$ and $H_{l|m}(p)=ph_{l|m}'(p)/h_{l|m}(p)$ for all $p\in(0,1)$. Then the following results hold.
\begin{itemize}
  \item [$(i)$] $H_{k|n}(p)$ is decreasing in $p\in(0,1);$ 
  \item [$(ii)$] ${H_{k|n}(p)}/{H_{l|m}(p)}$ is decreasing in $p\in(0,1)$, for all $k\leq l$ and $m-l\leq n-k$; 
  \item [$(iii)$] $(1-p){H_{k|n}'(p)}/{H_{k|n}(p)}$ is decreasing in $p\in (0,1)$.
  \end{itemize}
\end{lem}
\begin{lem}\label{l4}
Let $h_{k|n}(\cdot)$ and $h_{l|m}(\cdot)$ be the reliability functions of the $k$-out-of-$n$ and the $l$-out-of-$m$ systems with i.i.d. components, respectively, where $1\leq k\leq n$ and $1\leq l\leq m$. Further, let $R_{k|n}(p)=(1-p)h_{k|n}'(p)/(1-h_{k|n}(p))$ and $R_{l|m}(p)=(1-p)h_{l|m}'(p)/(1-h_{l|m}(p))$ for all $p\in(0,1)$. Then the following results hold.
\begin{itemize}
  \item [$(i)$] $R_{k|n}(p)$ is increasing in $p\in(0,1);$ 
  \item [$(ii)$] ${R_{k|n}(p)}/{R_{l|m}(p)}$ is increasing in $p\in(0,1)$, for all $l\leq k$ and $n-k\leq m-l$; 
  \item [$(iii)$] $p{R_{k|n}'(p)}/{R_{k|n}(p)}$ is decreasing in $p\in (0,1)$.
  \end{itemize}
\end{lem}
 \section{Stochastic comparisons of two coherent systems}\label{se3}
In this section we compare two coherent systems with respect to ageing faster orders in terms of the failure and the reversed failure rates. We show that the proposed results hold for the $k$-out-of-$n$ and the $l$-out-of-$m$ systems with i.i.d. components.
\\\hspace*{0.2 in}Let $\tau_1\left(\mbox{\boldmath$X$}\right)$ and $\tau_2\left(\mbox{\boldmath$Y$}\right)$ (resp. $\tau_{k|n}\left(\mbox{\boldmath$X$}\right)$ and $\tau_{l|m}\left(\mbox{\boldmath$Y$}\right)$) be the lifetimes of two coherent systems (resp. $k$-out-of-$n$ and $l$-out-of-$m$ systems) formed by two different sets of d.i.d. components with the lifetime vectors $\mbox{\boldmath$X$}=(X_1,X_2,\dots,X_{n})$ and $\mbox{\boldmath$Y$}=(Y_1,Y_2,\dots,Y_{m})$, respectively. For the sake of simplicity of notation, let us assume that all $X_i$'s are identically distributed with a non-negative random variable $X$, and all $Y_j$'s are identically distributed with a non-negaive random variable $Y$. Further, let $h_1(\cdot)$ and $h_2(\cdot)$ be the domination functions of $\tau_1\left(\mbox{\boldmath$X$}\right)$ and $\tau_2\left(\mbox{\boldmath$Y$}\right)$, respectively. In what follows, we use the following notation. For $p\in(0,1)$,
   $$H_i(p)=\frac{ph_i'(p)}{h_i(p)}, \quad i=1,2,$$ and $$R_i(p)=\frac{(1-p)h_i'(p)}{1-h_i(p)}, \quad i=1,2.$$ 
   \hspace*{0.2 in}In the following theorem we show that under a set of sufficient conditions $\tau_1\left(\mbox{\boldmath$X$}\right)$ is ageing faster than $\tau_2\left(\mbox{\boldmath$Y$}\right)$ in terms of the failure rate. 
 \begin{thm}\label{t1}
Suppose that the following conditions hold.
\begin{itemize}
\item [$(i)$] $H_1(p)$ and ${H_1(p)}/{H_2(p)}$ are decreasing in $p\in(0,1)$;
\item [$(ii)$] $(1-p){H_1'(p)}/{H_1(p)}$ or $(1-p){H_2'(p)}/{H_2(p)}$ is decreasing in $p\in (0,1)$;
\item [$(iii)$] $X\underset{ c}{\prec} Y$ and $Y\leq_{rh} X $.
\end{itemize}
Then $\tau_1\left(\mbox{\boldmath$X$}\right)\underset{ c}{\prec}\tau_2\left(\mbox{\boldmath$Y$}\right)$.$\hfill\Box$
\end{thm}
\hspace*{0.2 in} The following corollary follows from Theorem~\ref{t1} and Lemma~\ref{l3}. It is worthwile to mention here that Theorem~3.1 (a) of Misra and Francis~\cite{mf} is the particular case of this corollary ($k=l$ and $m=n$). 
  \begin{cor}\label{c1}
  Suppose that the $X_i$'s are i.i.d., and that the $Y_j$'s are i.i.d.
  If $X\underset{ c}{\prec} Y$ and $Y\leq_{rh} X $, then ${\tau_{k|n}\left(\mbox{\boldmath$X$}\right)}\underset{ c}{\prec}$ ${\tau_{l|m}\left(\mbox{\boldmath$Y$}\right)}$ for $k\leq l$ and $m-l\leq n-k$.
  \end{cor}
  \begin{rem}
 Let the assumption of Corollary~\ref{c1} hold. Then from Corollary~\ref{c1} we have
  \begin{itemize}
  \item [$(i)$] ${\tau_{k|n}\left(\mbox{\boldmath$X$}\right)}\underset{ c}{\prec}$ ${\tau_{l|n}\left(\mbox{\boldmath$Y$}\right)}$ for $k\leq l$;
  \item [$(ii)$] ${\tau_{k|n}\left(\mbox{\boldmath$X$}\right)}\underset{ c}{\prec}$ ${\tau_{k|m}\left(\mbox{\boldmath$Y$}\right)}$ for $m\leq n$;
  \item [$(iii)$] ${\tau_{l-r|m-r}\left(\mbox{\boldmath$X$}\right)}\underset{ c}{\prec}$ ${\tau_{l|m}\left(\mbox{\boldmath$Y$}\right)}$ for $r\leq l$.$\hfill\Box$
  \end{itemize}
  \end{rem}
  \hspace*{0.2 in}One natural question may arise, which is whether the result stated in Theorem~\ref{t1} holds without the condition $Y\leq_{rh} X$. Below we cite a counterexample which shows that this condition could not be relaxed.
  \begin{ce}
  Consider two coherent systems $\tau_{1}(\mbox{\boldmath$X$})=\max\{X_1,X_2,X_3\}$ and $\tau_{2}(\mbox{\boldmath$Y$})=\max\{Y_1,Y_2,Y_3\}$, where $X_i$'s are i.i.d. with the reliability function given by $\bar F_X(x)=\exp\{-2x^3\}$, $x>0$, and $Y_i$'s are i.i.d. with the reliability function given by $\bar F_Y(x)=\exp\{-0.1x^2\}$, $x>0$. Then it could easily be verified that $X\underset{ c}{\prec} Y$ but $Y\nleq_{rh} X $ (In fact $Y\nleq_{st} X $). Now, by writing $k(x)=r_{\tau_{1}\left(\mbox{\boldmath$X$}\right)}(x)/r_{\tau_{2}\left(\mbox{\boldmath$Y$}\right)}(x)$, we have
  \begin{eqnarray*}
 k(x)=30x e^{-(2x^3-0.1x^2)}\left[\frac{1-\left(1-e^{-0.1 x^2}\right)^3}{1-\left(1-e^{-2 x^3}\right)^3}\right]\left[\frac{\left(1-e^{-2 x^3}\right)^2}{\left(1-e^{-0.1 x^2}\right)^2}\right],\quad x>0,
  \end{eqnarray*}
  which is non-monotone over $x>0$, and hence ${\tau_{1}\left(\mbox{\boldmath$X$}\right)}\underset{ c}{\nprec}$ ${\tau_{2}\left(\mbox{\boldmath$Y$}\right)}$.$\hfill\Box$
  \end{ce}
  \hspace*{0.2 in}In the following proposition we give a necessary and sufficient condition for the case when the lifetimes of the components of both coherent systems are identically distributed. The proof follows in the same line as in Theorem~\ref{t1}, and hence omitted. 
   \begin{pro}\label{p1}
Let $X_i$'s be identically distributed. Then 
$\tau_1\left(\mbox{\boldmath$X$}\right)\underset{ c}{\prec}(\text{resp. }\underset{ c}{\succ})~\tau_2\left(\mbox{\boldmath$X$}\right)$ if and only if
\begin{eqnarray*}
{H_1(p)}/{H_2(p)}\text{ is decreasing (resp. increasing) in }p\in(0,1).
\end{eqnarray*}
\end{pro}
\hspace*{0.2 in}The following corollary, which is obtained in Theorem~2.1 of Misra and Francis~\cite{mf}, follows from Proposition~\ref{p1} and Lemma~\ref{l3}.
\begin{cor}\label{c2}
  Suppose that the $X_i$'s are i.i.d. Then ${\tau_{k|n}\left(\mbox{\boldmath$X$}\right)}\underset{ c}{\prec}$ ${\tau_{l|m}\left(\mbox{\boldmath$X$}\right)}$ for $k\leq l$ and $m-l\leq n-k$.
  \end{cor}
  \begin{rem}
   Let the assumption of Corollary~\ref{c2} hold. Then from Corollary~\ref{c2} we have
     \begin{itemize}
  \item [$(i)$] ${\tau_{k|n}\left(\mbox{\boldmath$X$}\right)}\underset{ c}{\prec}$ ${\tau_{l|n}\left(\mbox{\boldmath$X$}\right)}$ for $k\leq l$;
  \item [$(ii)$] ${\tau_{k|n}\left(\mbox{\boldmath$X$}\right)}\underset{ c}{\prec}$ ${\tau_{k|m}\left(\mbox{\boldmath$X$}\right)}$ for $m\leq n$;
  \item [$(iii)$] ${\tau_{l-r|m-r}\left(\mbox{\boldmath$X$}\right)}\underset{ c}{\prec}$ ${\tau_{l|m}\left(\mbox{\boldmath$X$}\right)}$ for $r\leq l$.$\hfill\Box$
  \end{itemize}
  \end{rem}
  \hspace*{0.2 in}The following corollary given in Ding and Zhang~\cite{dz} follows from Proposition~\ref{p1}. It shows that a series system ages faster (in terms of the hazard rate) as its number of components increases whereas the reverse scenario is observed for the parallel system.
  \begin{cor}
  Suppose that the $X_i$'s are d.i.d.components with the common Archimedean copula generated by $\phi(\cdot)$. If $x\ln'\left[-\phi'(x)/(1-\phi(x))\right]$ is decreasing in $x>0$, then
   \begin{itemize}
  \item [$(i)$] ${\tau_{1|n}\left(\mbox{\boldmath$X$}\right)}\underset{ c}{\prec}$ ${\tau_{1|m}\left(\mbox{\boldmath$X$}\right)}$ for $m\leq n$;
   \item [$(ii)$] ${\tau_{n|n}\left(\mbox{\boldmath$X$}\right)}\underset{ c}{\prec}$ ${\tau_{m|m}\left(\mbox{\boldmath$X$}\right)}$ for $n\leq m$. $\hfill\Box$
 \end{itemize}
  \end{cor}
\hspace*{0.2 in}Below we give an example that illustrates the result given in Proposition~\ref{p1}.
  \begin{ex}\label{exx1}
  Consider two coherent systems ${\tau_{1}\left(\mbox{\boldmath$X$}\right)}=\min\{X_1,\max\{X_2,X_3\}\}$ and ${\tau_{2}\left(\mbox{\boldmath$X$}\right)}=\min\{X_1,X_2,X_3\}$ which are formed by three identical components with lifetimes $X_1,X_2$ and $X_3$. Further, let the joint distribution function of $(X_1,X_2,X_3)$ be described by the FGM copula 
  $$K(p_1,p_2,p_3)=p_1p_2p_3(1+\theta (1-p_1)(1-p_2)(1-p_3)),$$ 
  where $p_i\in(0,1)$, $i=1,2,3$, and $\theta\in [-1,1]$.
  Then the domination functions of ${\tau_{1}\left(\mbox{\boldmath$X$}\right)}$ and ${\tau_{2}\left(\mbox{\boldmath$X$}\right)}$ are, respectively, given by
  \begin{eqnarray*}
  h_1(p)=2p^2-p^3-\theta p^3 (1-p)^3, \quad 0<p<1
  \end{eqnarray*}
  and
   \begin{eqnarray*}
  h_2(p)=p^3+\theta p^3(1-p)^3, \quad 0<p<1.
  \end{eqnarray*}
  These give
    \begin{eqnarray*}
 H_1(p)=\frac{ph_1'(p)}{h_1(p)}=\frac{4p^2-3(1+\theta)p^3+12\theta p^4-15\theta p^5+6\theta p^6}{2p^2-(1+\theta)p^3+3\theta p^4-3\theta p^5+\theta p^6}, \quad 0<p<1
  \end{eqnarray*}
  and 
  $$H_2(p)=\frac{ph_2'(p)}{h_2(p)}=\frac{3(1+\theta)p^3-6\theta p^6-12\theta p^4+15 \theta p^5}{(1+\theta)p^3-\theta p^6-3\theta p^4+3\theta p^5},\quad 0<p<1.$$
  Writing $s_\theta(p)=H_1(p)/H_2(p)$, we have
  $$s_\theta(p)=\frac{(4p^2-3(1+\theta)p^3+12\theta p^4-15\theta p^5+6\theta p^6)((1+\theta)p^3-\theta p^6-3\theta p^4+3\theta p^5)}{(2p^2-(1+\theta)p^3+3\theta p^4-3\theta p^5+\theta p^6)(3(1+\theta)p^3-6\theta p^6-12\theta p^4+15 \theta p^5)},\quad 0<p<1.$$
  Now, consider the following two cases.
\\Case-I: Let $\theta=-.9,-.8,\dots, .4,.5$. Then it can be verified that $s_\theta(p)=H_1(p)/H_2(p)$ is decreasing in $p\in(0,1)$, and hence $\tau_1\left(\mbox{\boldmath$X$}\right)\underset{ c}{\prec}\tau_2\left(\mbox{\boldmath$X$}\right)$ follows from Proposition~\ref{p1}.
\\Case-II: Let $\theta=.75,.80,\dots, .95,1$. Then it can be checked that $s_\theta(p)=H_1(p)/H_2(p)$ is non-monotone over $p\in(0,1)$. Hence, by Proposition~\ref{p1}, we get that neither $\tau_1\left(\mbox{\boldmath$X$}\right)\underset{ c}{\prec}\tau_2\left(\mbox{\boldmath$X$}\right)$ nor $\tau_1\left(\mbox{\boldmath$X$}\right)\underset{ c}{\succ}\tau_2\left(\mbox{\boldmath$X$}\right)$ holds.
$\hfill\Box$
 \end{ex}
  \hspace*{0.2 in}In the following theorem we compare $\tau_1\left(\mbox{\boldmath$X$}\right)$ and $\tau_2\left(\mbox{\boldmath$Y$}\right)$ with respect to the ageing faster order in the reversed hazard rate.
 \begin{thm}\label{t2}
Suppose that the following conditions hold.
\begin{itemize}
\item [$(i)$] $R_1(p)$ and ${R_1(p)}/{R_2(p)}$ are increasing in $p\in(0,1)$;
\item [$(ii)$] $p{R_1'(p)}/{R_1(p)}$ or $p{R_2'(p)}/{R_2(p)}$ is decreasing in $p\in (0,1)$;
\item [$(iii)$] $X\underset{ b}{\prec} Y$ and $X\leq_{hr} Y $.
\end{itemize}
Then $\tau_1\left(\mbox{\boldmath$X$}\right)\underset{ b}{\prec}\tau_2\left(\mbox{\boldmath$Y$}\right)$.$\hfill\Box$
\end{thm}
\hspace*{0.2 in} The following corollary immediately follows from Theorem~\ref{t2} and Lemma~\ref{l4}. Note that Theorem 3.1(b) of Misra and Francis~\cite{mf} is a particular case of this corollary ($k=l$ and $m=n$).
  \begin{cor}\label{c3}
  Suppose that the $X_i$'s are i.i.d., and that the $Y_j$'s are i.i.d.
  If $X\underset{ b}{\prec} Y$ and $X\leq_{hr} Y $, then ${\tau_{k|n}\left(\mbox{\boldmath$X$}\right)}\underset{ b}{\prec}$ ${\tau_{l|m}\left(\mbox{\boldmath$Y$}\right)}$ for $l\leq k$ and $ n-k\leq m-l$.
  \end{cor}
    \begin{rem}
   Let the assumption of Corollary~\ref{c3} hold. Then from Corollary~\ref{c3} we have
   \begin{itemize}
  \item [$(i)$] ${\tau_{k|n}\left(\mbox{\boldmath$X$}\right)}\underset{ b}{\prec}$ ${\tau_{l|n}\left(\mbox{\boldmath$Y$}\right)}$ for $l\leq k$;
  \item [$(ii)$] ${\tau_{k|n}\left(\mbox{\boldmath$X$}\right)}\underset{ b}{\prec}$ ${\tau_{k|m}\left(\mbox{\boldmath$Y$}\right)}$ for $n\leq m$;
  \item [$(iii)$] ${\tau_{k|n}\left(\mbox{\boldmath$X$}\right)}\underset{ b}{\prec}$ ${\tau_{k-r|n-r}\left(\mbox{\boldmath$Y$}\right)}$ for $r\leq k$. $\hfill\Box$
  \end{itemize}
   \end{rem}
    \hspace*{0.2 in}The following counterexample shows that the result given in Theorem~\ref{t2} may not hold without the condition $X\leq_{hr}Y.$
  \begin{ce}
  Consider the coherent systems $\tau_{1}(\mbox{\boldmath$X$})=\min\{X_1,X_2\}$ and $\tau_{2}(\mbox{\boldmath$Y$})=\min\{Y_1,Y_2\}$, where $X_i$'s are i.i.d. with the cumulative distribution function given by $ F_X(x)=\exp\{-(2.1/x)^7\}$, $x>0$, and $Y_i$'s are i.i.d. with the cumulative distribution function given by $ F_Y(x)=\exp\{-(2/x)^3\}$, $x>0$. Then it is easy to verify that $X\underset{ b}{\prec} Y$ but $X\nleq_{hr} Y $ (In fact $X\nleq_{st} Y $). Now, by writing $l(x)=\tilde r_{\tau_{2}\left(\mbox{\boldmath$Y$}\right)}(x)/\tilde r_{\tau_{1}\left(\mbox{\boldmath$X$}\right)}(x)$, we have
  \begin{eqnarray*}
  l(x)=\left[\frac{24 x^4 e^{-(2/x)^3}\left(1-e^{-(2/x)^3}\right)}{7\times 2.1^7e^{-(2.1/x)^7}\left(1-e^{-(2.1/x)^7}\right)}\right]\left[\frac{1-\left(1-e^{-(2.1/x)^7}\right)^2}{1-\left(1-e^{-(2/x)^3}\right)^2}\right],\quad x>0,
  \end{eqnarray*}
which is non-monotone over $x>0$, and hence ${\tau_{2|2}\left(\mbox{\boldmath$X$}\right)}\underset{ b}{\nprec}$ ${\tau_{2|2}\left(\mbox{\boldmath$Y$}\right)}$.$\hfill\Box$
  \end{ce}
\hspace*{0.2 in}In the following proposition we discuss an analog of Proposition~\ref{p1} for the ageing faster order in the reversed hazard rate.
    \begin{pro}\label{p2}
Let $X_i$'s be identically distributed. Then 
$\tau_1\left(\mbox{\boldmath$X$}\right)\underset{ b}{\prec}(\text{resp. }\underset{ b}{\succ})~\tau_2\left(\mbox{\boldmath$X$}\right)$ if and only if
$${R_1(p)}/{R_2(p)}\text{ is increasing (resp. decreasing) in }p\in(0,1).$$
\end{pro}
 \hspace*{0.2 in}The following corollary given in Theorem 2.2 of Misra and Francis~\cite{mf} follows from Proposition~\ref{p2} and Lemma~\ref{l4}.
 \begin{cor}\label{c4}
  Suppose that the $X_i$'s are i.i.d. Then ${\tau_{k|n}\left(\mbox{\boldmath$X$}\right)}\underset{ b}{\prec}$ ${\tau_{l|m}\left(\mbox{\boldmath$X$}\right)}$ for $l\leq k$ and $ n-k\leq m-l$.
  \end{cor}
    \begin{rem}
   Let the assumption of Corollary~\ref{c4} hold. Then from Corollary~\ref{c4} we have
    \begin{itemize}
  \item [$(i)$] ${\tau_{k|n}\left(\mbox{\boldmath$X$}\right)}\underset{ b}{\prec}$ ${\tau_{l|n}\left(\mbox{\boldmath$X$}\right)}$ for $l\leq k$;
  \item [$(ii)$] ${\tau_{k|n}\left(\mbox{\boldmath$X$}\right)}\underset{ b}{\prec}$ ${\tau_{k|m}\left(\mbox{\boldmath$X$}\right)}$ for $n\leq m$;
  \item [$(iii)$] ${\tau_{k|n}\left(\mbox{\boldmath$X$}\right)}\underset{ b}{\prec}$ ${\tau_{k-r|n-r}\left(\mbox{\boldmath$X$}\right)}$ for $r\leq k$.$\hfill\Box$
  \end{itemize}
  \end{rem}
  \hspace*{0.2 in}The following corollary obtained in Ding and Zhang~\cite{dz} immediately follows from Proposition~\ref{p2}. It shows that a series system ages faster (in terms of the reversed hazard rate) as its number of components decreases whereas the reverse scenario is observed for the parallel system. 
  \begin{cor}
  Suppose that the $X_i$'s are d.i.d. components with the common Archimedean copula generated by $\phi(\cdot)$. If $x\ln'\left[-\phi'(x)/\phi(x)\right]$ is decreasing (resp. increasing) in $x>0$, then
   \begin{itemize}
  \item [$(i)$] ${\tau_{1|n}\left(\mbox{\boldmath$X$}\right)}\underset{ b}{\succ}(\text{resp. }\underset{ b}{\prec})$ ${\tau_{1|m}\left(\mbox{\boldmath$X$}\right)}$ for $m\leq n$;
   \item [$(ii)$] ${\tau_{n|n}\left(\mbox{\boldmath$X$}\right)}\underset{ b}{\succ}(\text{resp. }\underset{ b}{\prec})$ ${\tau_{m|m}\left(\mbox{\boldmath$X$}\right)}$ for $n\leq m$. $\hfill\Box$
 \end{itemize}
  \end{cor}
 \hspace*{0.2 in}The result stated in Proposition~\ref{p2} is revealed through the following example.  
   \begin{ex}\label{exx2}
  Consider two coherent systems which are discussed in Example~\ref{exx1}. Then
    \begin{eqnarray*}
 R_1(p)=\frac{(1-p)h_1'(p)}{1-h_1(p)}=\frac{4p-(7+3\theta)p^2+3(1+5\theta)p^3-27\theta p^4+21\theta p^5-6\theta p^6}{1-2p^2+(1+\theta)p^3-3\theta p^4+3\theta p^5-\theta p^6},\quad 0<p<1
  \end{eqnarray*}
  and 
  $$R_2(p)=\frac{(1-p)h_2'(p)}{1-h_2(p)}=\frac{3(1+\theta)p^2-3(1+5\theta)p^3+27\theta p^4-21\theta p^5+6\theta p^6}{1-(1+\theta)p^3+\theta p^6+3\theta p^4-3\theta p^5},\quad 0<p<1.$$
  Writing $v_\theta(p)=R_2(p)/R_1(p)$, we have
  \begin{eqnarray*}
  v_\theta(p)=&&\frac{3(1+\theta)p^2-3(1+5\theta)p^3+27\theta p^4-21\theta p^5+6\theta p^6}{1-(1+\theta)p^3+\theta p^6+3\theta p^4-3\theta p^5}
  \\&&\times \frac{1-2p^2+(1+\theta)p^3-3\theta p^4+3\theta p^5-\theta p^6}{4p-(7+3\theta)p^2+3(1+5\theta)p^3-27\theta p^4+21\theta p^5-6\theta p^6},\quad 0<p<1.
  \end{eqnarray*}
 For $\theta=-1,-.8,\dots, .8,1$, it can be verified that $v_\theta(p)$ is increasing in $p\in(0,1)$. Hence $\tau_1\left(\mbox{\boldmath$X$}\right)\underset{ b}{\succ}\tau_2\left(\mbox{\boldmath$X$}\right)$ follows from Proposition~\ref{p2}.
 \end{ex}
  \section{Stochastic comparisons of coherent systems with active redundancy at the component level versus the system level}\label{se4}
 Let ${\mbox{\boldmath $X$}}=(X_1,X_2,\dots,X_n)$ be a vector of random variables representing the lifetimes of $n$ d.i.d. components. Further, let $\{{\mbox{\boldmath $Y$}}_1,{\mbox{\boldmath $Y$}}_2,\dots,{\mbox{\boldmath $Y$}}_m\}$ be a set of $m$ vectors representing the lifetimes of $mn$ spares (or redundant components), where ${\mbox{\boldmath $Y$}}_i=(Y_{i1},Y_{i2},\dots,Y_{in})$ is a vector of $n$ d.i.d. random variables, for $i=1,2,\dots,m$. Assume that all $X_j$'s and $Y_{ij}$'s are identically distributed with a non-negaive random variable $X$. We write $T_C=\tau\left({\mbox{\boldmath $X \vee Y_1\vee  Y_2\vee  \dots \vee  Y_m$}}\right)$ to denote the lifetime of a coherent system with active redundancies at the component level, where the symbol ${\mbox{\boldmath $X \vee Y_1\vee  Y_2\vee  \dots \vee  Y_m$}}$ stands for a $n$-tuple vector ${\mbox{\boldmath $Z$}}=(Z_1,Z_2,\dots,Z_n)$ such that $Z_j$ represents the lifetime of a parallel system formed by $(m+1)$ independent components $\{X_j,Y_{1j},\dots,Y_{mj}\}$, for $j=1,2,\dots,n$. Further, we write
$T_S=\tau({\mbox{\boldmath $X$}})\vee  \tau({\mbox{\boldmath $Y$}}_1)\vee  \tau({\mbox{\boldmath $Y$}}_2)\vee  \dots \vee \tau({\mbox{\boldmath $Y$}}_m)$ to denote the lifetime of a coherent system with active redundancies at the system level, where the symbol $\vee$ stands for maximum. Furthermore, it is assumed that $\tau({\mbox{\boldmath $X$}})$ and $ \tau({\mbox{\boldmath $Y$}}_i)$'s are independent, and they have the same domination function as $\tau({\mbox{\boldmath $Z$}})$ has. We denote this domination function by $h(\cdot)$. In what follows, we use the notation $R(p)=(1-p)h'(p)/(1-h(p))$, $p\in(0,1)$.
  \\\hspace*{0.2 in}In the following theorem, we provide an equivalent condition to hold that the allocation of redundancy at the component level is better/worse than that at the system level with respect to the ageing faster order in terms of the hazard rate. 
  \begin{thm}\label{t41}
  For $m\in \mathbb{N}$, 
$T_S \underset{ c}{\prec}(\text{resp. }\underset{ c}{\succ})~ T_C$ holds if and only if
 \begin{eqnarray}\label{km0}
\left(\frac{\left(1-h(p)\right)^{m}h'\left(p\right)}{1-\left(1-h(p)\right)^{m+1}}\right)\left(\frac{h\left(1-(1-p)^{m+1}\right)}{\left(1-p\right)^{m}h'\left(1-(1-p)^{m+1}\right)}\right)
 \end{eqnarray}
 $\text{ is decreasing (resp. increasing) in }p\in(0,1).$ $\hfill\Box$
  \end{thm}
\hspace*{0.2 in}The following corollary follows from Theorem~\ref{km0}. 
  \begin{cor}\label{co1}
  If all $X_i$'s and $Y_j$'s are i.i.d., then  $\tau_{n|n}\left(\mbox{\boldmath$X$}\right)\vee \tau_{n|n}\left(\mbox{\boldmath$Y_1$}\right)\underset{ c}{\succ}\tau_{n|n}\left(\mbox{\boldmath$X\vee Y_1$}\right).$ $\hfill\Box$
  \end{cor}
 \hspace*{0.2 in}In the next theorem we discuss an analog of Theorem~\ref{t41} under the ageing faster order in the reversed hazard rate.
 \begin{thm}\label{t3}
 For $m\in \mathbb{N}$, 
$T_S \underset{ b}{\prec} T_C$ holds if and only if
 \begin{eqnarray*}\label{cs1}
 ~~~~~~~~~~~~~~~~~~~~~~~~~
 \frac{R(p)}{R(1-(1-p)^{m+1})}\text{ is increasing in }p\in(0,1).
 ~~~~~~~~~~~~~~~~~~~~~~~~~~~~~~~~~\hfill\Box
 \end{eqnarray*}
 \end{thm}
\hspace*{0.2 in}Since the condition given in Theorem~\ref{t3} is involved with $m$, it is practically not easy to verify. In the following proposition we discuss a sufficient condition that could be useful to show the result.
  \begin{pro}\label{p41}
  If $pR'(p)/R(p)$ is decreasing and positive for all $p\in(0,1)$, then $T_S \underset{ b}{\prec} T_C$.
  \end{pro}
\hspace*{0.2 in}The following corollary follows from Proposition~\ref{p41} and Lemma~\ref{l4}.
 \begin{cor}
 Suppose that all $X_i$'s and $Y_j$'s are i.i.d. Then, for $1\leq k\leq n,$ $$\tau_{k|n}\left(\mbox{\boldmath$X$}\right)\vee \tau_{k|n}\left(\mbox{\boldmath$Y_1$}\right)\vee \dots \vee\tau_{k|n}\left(\mbox{\boldmath$Y_m$}\right)\underset{ b}{\prec}\tau_{k|n}\left(\mbox{\boldmath$X\vee Y_1 \vee\dots \vee Y_m$}\right).$$
 \end{cor}
 \hspace*{0.2 in}Below we provide an example that illustrates the result given in Proposition~\ref{p41}.
   \begin{ex}\label{exx3}
  Let $m=1$. Consider a coherent system ${\tau\left(\mbox{\boldmath$X$}\right)}=\min\{X_1,X_2,\dots,X_{n}\}$ formed by $n$ identical components with the lifetime vector $\mbox{\boldmath$X$}=(X_1,X_2,\dots,X_{n})$. Further, let $\{X_1,X_2,\dots,X_n\}$ have the Gumbel-Hougard copula given by
  $$K(p_1,p_2,\dots,p_{n})=\exp\left\{-\left(\sum\limits_{i=1}^{n}(-\ln p_i)^\theta\right)^{1/\theta}\right\},$$ 
  where $p_i\in(0,1)$, $i=1,2,\dots,n$, and $\theta\in [1,\infty)$.
  Then the domination function of ${\tau\left(\mbox{\boldmath$X$}\right)}$ is given by
  $h(p)=p^{a}$, 
  where $a=n^{1/\theta}$ ($\geq 1$). This gives 
  $$R(p)=\frac{(1-p)h'(p)}{(1-h(p))}=\frac{a\left(p^{a-1}-p^a\right)}{1-p^a},\quad 0<p<1$$
  and
  $$\frac{pR'(p)}{R(p)}=\frac{a-1-ap+p^a}{1-p-p^a+p^{a+1}},\quad 0<p<1.$$
  Since $((1-p^a)/(1-p))\leq a$, for all $a\geq 1$ and $p\in(0,1)$, we have $pR'(p)/R(p)\geq 0$, for all $p\in(0,1)$. Further,
  \begin{eqnarray*}
  \left[\frac{pR'(p)}{R(p)}\right]'=\frac{\gamma_1(p)}{\left(1-p-p^a+p^{a+1}\right)^2}, \quad 0<p<1,
  \end{eqnarray*}
  where $$\gamma_1(p)=a^2p^{a-1}+2(1-a^2)p^a+a^2 p^{a+1}-p^{2a}-1,\quad 0<p<1.$$ 
  \end{ex}
  Now,
  $${\gamma_1}'(p)=p^{a-2}\gamma_2(p),\quad 0<p<1,$$
  where $$\gamma_2(p)=a^2(a-1)-2a(a^2-1)p+a^2(a+1)p^2-2ap^{a+1},\quad 0<p<1.$$
  Differentiating $\gamma_2(p)$ twice, we get
  \begin{eqnarray*}
  &&\gamma_2'(p)=-2a(a^2-1)+2a^2(a+1)p-2a(a+1)p^a,\quad 0<p<1
   \end{eqnarray*}
   and
   \begin{eqnarray*}
  \gamma_2''(p)=2a^2(a+1)\left(1-p^{a-1}\right)\geq 0,\quad 0<p<1.
  \end{eqnarray*}
 Thus, we have $\gamma_2'(p)\leq \gamma_2'(1)=0$, for all $p\in(0,1)$, which implies $\gamma_2(p)\geq \gamma_2(1)=0$, for all $p\in(0,1)$. Further, this implies $\gamma_1'(p)\geq 0$, for all $p\in(0,1)$, which gives $\gamma_1(p)\leq \gamma_1(1)=0$, for all $p\in(0,1)$.
   Hence $pR'(p)/R(p)$ is decreasing in $p\in(0,1)$.
   Thus, $T_S \underset{ b}{\prec} T_C$ follows from Proposition~\ref{p41}.
 \section{Stochastic comparisons of a used coherent system and a coherent system of used components}\label{se5}
Let $X$ be a random variable representing the lifetime of a component/system. Then its residual lifetime at a time instant $t$ ($>0$) is denoted by $X_t$ and is defined as
 $$X_t=(X-t|X>t).$$
 We call $X_t$ as a used component/system. Let $\mbox{\boldmath$X$}=(X_1,X_2,\dots,X_n)$ be a vector of random variables representing the lifetimes of $n$ d.i.d. components. Then we write
 $$\mbox{\boldmath$X$}_t=\left((X_1)_t,(X_2)_t,\dots,(X_n)_t\right),\quad t>0,$$
 to represent a vector of $n$ used components $\{(X_1)_t,(X_2)_t,\dots,(X_n)_t\}$, $t>0$. Consequently, we write $\tau\left(\mathbf{X}_{t}\right)$ to denote the lifetime of a coherent system made by a set of components with the lifetime vector $\mbox{\boldmath$X$}_t$. Further, by $\left(\tau\left(\mathbf{X}\right)\right)_{t}=(\tau\left(\mathbf{X}\right)-t|\tau\left(\mathbf{X}\right)>t)$, we mean the lifetime of a used coherent system formed by a set of components with the lifetime vector $\mbox{\boldmath$X$}$. For the sake of simplicity, we assume that all $X_i$'s are identically distributed with a non-negative random variable $X$. In what follows, we denote the reliability function of $\tau\left(\mathbf{X}\right)$ by $h(\cdot)$, and we write $H(p)=ph'(p)/h(p)$, $0<p<1$.
 \\\hspace*{0.2 in}In the following theorem we derive the necessary and sufficient condition for a used coherent system to be ageing faster than a coherent system of used components in terms of the hazard rate.
  \begin{thm}\label{th1}
For any fixed $t\geq0$, $\tau\left(\mathbf{X}_{t}\right)\underset{ c}{\prec}\left(\tau\left(\mathbf{X}\right)\right)_{t}$ holds if and only if
\begin{eqnarray}\label{rs2}
{pH'(p)}/{H(p)}\text{ is decreasing in }p\in(0,1).
\end{eqnarray}
\end{thm}
\hspace*{0.2 in}In the following proposition we discuss the same result as in Theorem~\ref{th1} under a different set of sufficient conditions which is sometimes easy to verify. The proof follows from Theorem~\ref{th1}. Hence we omit it.  
   \begin{pro}\label{pp1}
For any fixed $t\geq 0$, 
$\tau\left(\mbox{\boldmath$X$}_t\right)\underset{ c}{\prec}\left(\tau\left(\mbox{\boldmath$X$}\right)\right)_t$ holds if
\begin{eqnarray*}
 (1-p){H'(p)}/{H(p)} \text{ is decreasing and negative in }p\in(0,1).
\end{eqnarray*}
\end{pro}
\hspace*{0.2 in}The following corollary follows from Proposition~\ref{pp1} and Lemma~\ref{l3}.
\begin{cor}
If the $X_i$'s are i.i.d., then ${\tau_{k|n}\left(\mbox{\boldmath$X$}_t\right)}\underset{ c}{\prec}\left({\tau_{k|n}\left(\mbox{\boldmath$X$}\right)}\right)_t$, for any fixed $t\geq 0$, and $1\leq k\leq n$.$\hfill\Box$
  \end{cor}
\hspace*{0.2 in}In the following theorem we show a similar result as in Theorem~\ref{th1} for the ageing faster order in the reversed hazard rate.  
   \begin{thm}\label{th2}
For any fixed $t\geq 0$,  
$\tau\left(\mbox{\boldmath$X$}_t\right)\underset{ b}{\prec}(\text{resp. }\underset{ b}{\succ})\;\left(\tau\left(\mbox{\boldmath$X$}\right)\right)_t$ holds if and only if, for all $q\in(0,1)$,
\begin{eqnarray}\label{rs0}
\left[\frac{h'(p/q)}{h'(p)}\right]\left[\frac{h(q)-h(p)}{1-h(p/q)}\right]\text{ is increasing (resp. decreasing) in }p\in(0,q).
\end{eqnarray}
\end{thm}
\hspace*{0.2 in}As a consequence of Theorem~\ref{th2}, we have the following corollary. 
\begin{cor}\label{c02}
If the $X_i$'s are i.i.d., then $\tau_{1|n}\left(\mbox{\boldmath$X$}_t\right)\underset{ b}{\prec}\;\left(\tau_{1|n}\left(\mbox{\boldmath$X$}\right)\right)_t$, for any fixed $t\geq 0$.
\end{cor}
 \section{Concluding Remarks}\label{se6}
 In this paper, we study ageing faster orders (in terms of the hazard and the reversed hazard rates) which are useful to compare the relative ageings of two  systems. To be more specific, we provide sufficient conditions under which one coherent system is ageing more faster than another one with respect to the hazard and the reversed hazard rates. Further, we consider a problem of allocation of redundancies into a coherent system. We show that, under some necessary and sufficient conditions, the allocation of active redundancy at the component level is superior (inferior) to that at the system level with respect to ageing faster orders, for a coherent system. Furthermore, a used coherent system and a coherent system made out of used components are compared with respect to these ageing faster orders. 
 Apart from these, we also show that most of our developed results hold for the well known $k$-out-of-$n$ and the $l$-out-of-$m$ systems. Nevertheless, we provide a list of examples to illustrate our proposed results. Some counterexamples are also given wherever needed.
 \\\hspace*{0.2 in}Even though a vast literature exists on the study of different stochastic orders, there are a few results developed for the ageing faster orders. Since the ageing faster orders compare the relative ageings of two systems and the ageing is a common phenomenon experienced by each and every system, the study of ageing faster orders should be paid more attention from the researchers across the world. We believe that our study not only enriches the literature on ageing faster orders but also may be useful in some practical scenarios.
 \\\hspace*{0.2 in}Similar to the problems considered in this paper, the study of other stochastic orders (as discussed in the introduction section), which describe the relative ageings of two systems, is under investigation, and will be reported in future.
\subsection*{Acknowledgments}
\hspace*{0.2 in}The authors are thankful to the Editor-in-Chief, the Associate Editor and the anonymous Reviewers for their valuable constructive comments/suggestions which lead to an improved version of the manuscript. The first author sincerely acknowledges the financial support from the IIT Jodhpur, Karwar-$342037$, India.
  
{\bf Appendix}
\\\\{\bf Proof of Lemma~\ref{l3}($ii$):} Note that, for all $p\in(0,1),$
 \begin{eqnarray}\label{eq00}
  h_{k|n}(p)&=&\frac{1}{B(k,n-k+1)}\int\limits_0^p u^{k-1}(1-u)^{n-k}du,
 \end{eqnarray}
 where $B(\cdot,\cdot)$ is the the beta function. Then
 \begin{eqnarray}\label{eq0}
 \frac{1}{H_{k|n}(p)}=\int\limits_0^1 u^{k-1}\left(\frac{1-up}{1-p}\right)^{n-k}du, \quad 0<p<1
 \end{eqnarray}
  and
 \begin{eqnarray*}
 \frac{1}{H_{l|m}(p)}=\int\limits_0^1 u^{l-1}\left(\frac{1-up}{1-p}\right)^{m-l}du,\quad 0<p<1.
 \end{eqnarray*}
 Combing these two, we have
 \begin{eqnarray*}
 \frac{H_{l|m}(p)}{H_{k|n}(p)}=\frac{\int\limits_0^1 u^{k-1}\left(\frac{1-up}{1-p}\right)^{n-k}du}{\int\limits_0^1 u^{l-1}\left(\frac{1-up}{1-p}\right)^{m-l}du},\quad 0<p<1.
 \end{eqnarray*}
 Let $c$ be any real number. Consider the relation
 \begin{eqnarray*}
 {H_{l|m}(p)}-c{H_{k|n}(p)}\stackrel{sgn}=\int\limits_0^1 \xi_1(u,p)\eta_1(u,p)du,\quad 0<p<1,
 \end{eqnarray*}
 where 
 \begin{eqnarray*}
  \xi_1(u,p)=u^{l-1}\left(\frac{1-up}{1-p}\right)^{m-l},\quad 0<u<1,\;0<p<1
  \end{eqnarray*}
  and 
  \begin{eqnarray*}
  \eta_1(u,p)=u^{k-l}\left(\frac{1-up}{1-p}\right)^{n-k-m+l}-c,\quad 0<u<1,\;0<p<1.
 \end{eqnarray*}
 Note that
 \begin{eqnarray}\label{eq7}
 \xi_1(u,p) \text{ is RR}_2 \text{ in }(u,p)\in (0,1)\times (0,1)
 \end{eqnarray}
 and 
 \begin{eqnarray}\label{eq8}
 \eta_1(u,p) \text{ is increasing in } p\in (0,1), \text{ for all }u\in(0,1).
 \end{eqnarray}
 Further, since $k\leq l$ and $ m-l\leq n-k$, 
 \begin{eqnarray*}
{u^{k-l}\left(\frac{1-up}{1-p}\right)^{n-k-m+l}} \text{ is decreasing in } u \in (0,1),
 \end{eqnarray*}
for all $p\in(0,1)$. Then, on using Lemma~\ref{l2} we have that, for all $p\in(0,1)$, 
 $\eta_1(u,p)$ changes sign at most once, and if the change of sign does occur, it is from positive to negative, as $u$ traverses from $0$ to $1$. Finally, on using this together with \eqref{eq7} and \eqref{eq8} in  Lemma~\ref{l1}, we get that ${H_{l|m}(p)}-c{H_{k|n}(p)}$ changes sign at most once, and if the change of sign does occur, it is from negative to positive, as $u$ traverses from $0$ to $1$. Thus, on using Lemma~\ref{l2}, we get that ${H_{l|m}(p)}/{H_{k|n}(p)}$ is increasing in $p\in(0,1),$ which further implies that ${H_{k|n}(p)}/{H_{l|m}(p)}$ is decreasing in $p\in(0,1)$. Hence, the result is proved. $\hfill\Box$
\\\\{\bf Proof of Lemma~\ref{l3}($iii$):} Differentiating \eqref{eq0} on both sides, we get
 \begin{eqnarray*}
 -\frac{H'_{k|n}(p)}{\left[H_{k|n}(p)\right]^2}=\frac{n-k}{1-p}\int\limits_0^1 u^{k-1}\left(\frac{1-up}{1-p}\right)^{n-k-1}\left(\frac{1-u}{1-p}\right)du,\quad 0<p<1,
 \end{eqnarray*}
 which gives 
 \begin{eqnarray*}
 (1-p)\frac{H'_{k|n}(p)}{H_{k|n}(p)}=-\frac{(n-k)\int\limits_0^1 u^{k-1}\left(\frac{1-up}{1-p}\right)^{n-k-1}\left(\frac{1-u}{1-p}\right)du}{\int\limits_0^1 u^{k-1}\left(\frac{1-up}{1-p}\right)^{n-k}du},\quad 0<p<1.
 \end{eqnarray*}
 Thus, to prove the result it suffices to show that
 \begin{eqnarray}\label{eq9}
 \frac{N_1(p)}{D_1(p)}\stackrel{\text{def.}}=\frac{\int\limits_0^1 u^{k-1}\left(\frac{1-up}{1-p}\right)^{n-k-1}\left(\frac{1-u}{1-p}\right)du}{\int\limits_0^1 u^{k-1}\left(\frac{1-up}{1-p}\right)^{n-k}du}\text{ is increasing in }p\in(0,1).
 \end{eqnarray}
  Let $\alpha$ be any real number. Consider the relation
  \begin{eqnarray*}
  N_1(p)-\alpha D_1(p)\stackrel{sgn}=\int\limits_0^1 \xi_2(u,p)\eta_2(u,p)du,\quad 0<p<1,
 \end{eqnarray*}
 where 
 \begin{eqnarray*}
  \xi_2(u,p)=u^{k-1}\left(\frac{1-up}{1-p}\right)^{n-k},\quad 0<u<1,\;0<p<1
  \end{eqnarray*}
  and 
  \begin{eqnarray*}
  \eta_2(u,p)=\left(\frac{1-u}{1-up}\right)-\alpha,\quad 0<u<1,\;0<p<1.
 \end{eqnarray*}
  Note that
 \begin{eqnarray}\label{eq07}
 \xi_2(u,p) \text{ is RR}_2 \text{ in }(u,p)\in (0,1)\times (0,1)
 \end{eqnarray}
 and 
 \begin{eqnarray}\label{eq08}
 \eta_2(u,p) \text{ is increasing in } p\in (0,1), \text{ for all }u\in(0,1).
 \end{eqnarray}
 Further, it could be verified that, for all $p\in(0,1),$
 \begin{eqnarray*}
{\left(\frac{1-u}{1-up}\right)} \text{ is decreasing in } u \in (0,1).
 \end{eqnarray*}
Then, on using Lemma~\ref{l2} we have that 
 $\eta_2(u,p)$ changes sign at most once, and if the change of sign does occur, it is from positive to negative, as $u$ traverses from $0$ to $1$. Finally, on using this together with \eqref{eq07} and \eqref{eq08} in  Lemma~\ref{l1}, we get that $N_1(p)-\alpha D_1(p)$ changes sign at most once, and if the change of sign does occur, it is from negative to positive, as $u$ traverses from $0$ to $1$. Thus, on using Lemma~\ref{l2}, we get that $N_1(p)/ D_1(p)$ is increasing in $p\in(0,1),$ and hence the result is proved. $\hfill\Box$
\\\\{\bf Proof of Lemma~\ref{l4}($ii$):} From \eqref{eq00}, we have
 \begin{eqnarray}\label{eq20}
 \frac{1}{R_{k|n}(p)}=\int\limits_0^1 u^{n-k}\left(\frac{1-u(1-p)}{p}\right)^{k-1}du,\quad 0<p<1
 \end{eqnarray}
  and
 \begin{eqnarray*}
 \frac{1}{R_{l|m}(p)}=\int\limits_0^1 u^{m-l}\left(\frac{1-u(1-p)}{p}\right)^{l-1}du,\quad 0<p<1.
 \end{eqnarray*}
 Combing these two, we have
 \begin{eqnarray*}
 \frac{R_{l|m}(p)}{R_{k|n}(p)}=\frac{\int\limits_0^1 u^{n-k}\left(\frac{1-u(1-p)}{p}\right)^{k-1}du}{\int\limits_0^1 u^{m-l}\left(\frac{1-u(1-p)}{p}\right)^{l-1}du},\quad 0<p<1.
 \end{eqnarray*}
 Let $\beta$ be any real number. Consider the relation
 \begin{eqnarray*}
 {R_{l|m}(p)}-\beta{R_{k|n}(p)}\stackrel{sgn}=\int\limits_0^1 \xi_3(u,p)\eta_3(u,p)du,\quad 0<p<1,
 \end{eqnarray*}
 where 
 \begin{eqnarray*}
  \xi_3(u,p)=u^{m-l}\left(\frac{1-u(1-p)}{p}\right)^{l-1},\quad 0<u<1,\;0<p<1
  \end{eqnarray*}
  and 
  \begin{eqnarray*}
  \eta_3(u,p)= u^{n-k-m+l}\left(\frac{1-u(1-p)}{p}\right)^{k-l}-\beta,\quad 0<u<1,\;0<p<1.
 \end{eqnarray*}
 Note that
 \begin{eqnarray}\label{eq27}
 \xi_3(u,p) \text{ is TP}_2 \text{ in }(u,p)\in (0,1)\times (0,1)
 \end{eqnarray}
 and 
 \begin{eqnarray}\label{eq28}
 \eta_3(u,p) \text{ is decreasing in } p\in (0,1), \text{ for all }u\in(0,1).
 \end{eqnarray}
 Further, for all $p\in(0,1),$
 \begin{eqnarray*}
 u^{n-k-m+l}\left(\frac{1-u(1-p)}{p}\right)^{k-l} \text{ is decreasing in } u \in (0,1).
 \end{eqnarray*}
Then, on using Lemma~\ref{l2} we have that 
 $\eta_3(u,p)$ changes sign at most once, and if the change of sign does occur, it is from positive to negative, as $u$ traverses from $0$ to $1$. Finally, on using this together with \eqref{eq27} and \eqref{eq28} in  Lemma~\ref{l1}, we get that ${R_{l|m}(p)}-\beta{R_{k|n}(p)}$ changes sign at most once, and if the change of sign does occur, it is from positive to negative, as $u$ traverses from $0$ to $1$. Thus, on using Lemma~\ref{l2}, we get that ${R_{l|m}(p)}/{R_{k|n}(p)}$ is decreasing in $p\in(0,1),$ which further implies that ${R_{k|n}(p)}/{R_{l|m}(p)}$ is increasing in $p\in(0,1)$. Hence, the result is proved. $\hfill\Box$
\\\\{\bf Proof of Lemma~\ref{l4}($iii$):} Differentiating \eqref{eq20} on both sides, we get
 \begin{eqnarray*}
 \frac{R'_{k|n}(p)}{\left[R_{k|n}(p)\right]^2}=\frac{k-1}{p}\int\limits_0^1 u^{n-k}\left(\frac{1-u(1-p)}{p}\right)^{k-2}\left(\frac{1-u}{p}\right)du,\quad 0<p<1,
 \end{eqnarray*}
 which gives 
 \begin{eqnarray*}
 p\frac{R'_{k|n}(p)}{R_{k|n}(p)}=\frac{(k-1)\int\limits_0^1 u^{n-k}\left(\frac{1-u(1-p)}{p}\right)^{k-2}\left(\frac{1-u}{p}\right)du}{\int\limits_0^1 u^{n-k}\left(\frac{1-u(1-p)}{p}\right)^{k-1}du},\quad 0<p<1.
 \end{eqnarray*}
 Thus, to prove the result it suffices to show that
 \begin{eqnarray*}
 \frac{N_2(p)}{D_2(p)}\stackrel{\text{def.}}=\frac{\int\limits_0^1 u^{n-k}\left(\frac{1-u(1-p)}{p}\right)^{k-2}\left(\frac{1-u}{p}\right)du}{\int\limits_0^1 u^{n-k}\left(\frac{1-u(1-p)}{p}\right)^{k-1}du}\text{ is decreasing in }p\in(0,1).
 \end{eqnarray*}
  Let $\gamma$ be any real number. Consider the relation
  \begin{eqnarray*}
  N_2(p)-\gamma D_2(p)\stackrel{sgn}=\int\limits_0^1 \xi_4(u,p)\eta_4(u,p)du,\quad 0<p<1,
 \end{eqnarray*}
 where 
 \begin{eqnarray*}
  \xi_4(u,p)=u^{n-k}\left(\frac{1-u(1-p)}{p}\right)^{k-1},\quad 0<u<1,\;0<p<1
  \end{eqnarray*}
  and 
  \begin{eqnarray*}
  \eta_4(u,p)=\left(\frac{1-u}{1-u(1-p)}\right)-\gamma,\quad 0<u<1,\;0<p<1.
 \end{eqnarray*}
  Note that
 \begin{eqnarray}\label{eq007}
 \xi_4(u,p) \text{ is TP}_2 \text{ in }(u,p)\in (0,1)\times (0,1)
 \end{eqnarray}
 and 
 \begin{eqnarray}\label{eq008}
 \eta_4(u,p) \text{ is decreasing in } p\in (0,1), \text{ for all }u\in(0,1).
 \end{eqnarray}
 Further, it could be verified that, for all $p\in(0,1),$
 \begin{eqnarray*}
\frac{1-u}{1-u(1-p)} \text{ is decreasing in } u \in (0,1).
 \end{eqnarray*}
Then, on using Lemma~\ref{l2} we have that 
 $\eta_4(u,p)$ changes sign at most once, and if the change of sign does occur, it is from positive to negative, as $u$ traverses from $0$ to $1$. Finally, on using this together with \eqref{eq007} and \eqref{eq008} in  Lemma~\ref{l1}, we get that $N_2(p)-\alpha D_2(p)$ changes sign at most once, and if the change of sign does occur, it is from positive to negative, as $u$ traverses from $0$ to $1$. Thus, on using Lemma~\ref{l2}, we get that $N_2(p)/ D_2(p)$ is decreasing in $p\in(0,1),$ and hence the result is proved. $\hfill\Box$
\\\\{\bf Proof of Theorem~\ref{t1}:} Note that
\begin{eqnarray*}
\bar F_{\tau_1\left(\mbox{\boldmath$X$}\right)}(x)=h_1\left(\bar F_X(x)\right)\text{ and }\bar F_{\tau_2\left(\mbox{\boldmath$Y$}\right)}(x)=h_2\left(\bar F_Y(x)\right),\quad x>0,
\end{eqnarray*}
which gives failure rates of $\tau_1\left(\mbox{\boldmath$X$}\right)$ and $\tau_2\left(\mbox{\boldmath$Y$}\right)$ as
\begin{eqnarray*}
r_{\tau_1\left(\mbox{\boldmath$X$}\right)}(x)=\frac{f_X(x)h_1'\left(\bar F_X(x)\right)}{h_1\left(\bar F_X(x)\right)}=r_X(x)H_1(\bar F_X(x)),\quad x>0
\end{eqnarray*}
and
\begin{eqnarray*}
r_{\tau_2\left(\mbox{\boldmath$Y$}\right)}(x)=\frac{f_Y(x)h_2'\left(\bar F_Y(x)\right)}{h_2\left(\bar F_X(x)\right)}=r_Y(x)H_2(\bar F_Y(x)),\quad x>0,
\end{eqnarray*}
respectively.
 Then, $\tau_1\left(\mbox{\boldmath$X$}\right)\underset{ c}{\prec}\tau_2\left(\mbox{\boldmath$Y$}\right)$ holds if and only if
\begin{eqnarray*}
\frac{r_{\tau_1\left(\mbox{\boldmath$X$}\right)}(x)}{r_{\tau_2\left(\mbox{\boldmath$X$}\right)}(x)}=\left[\frac{r_X(x)}{r_Y(x)}\right]\left[\frac{H_1(\bar F_X(x))}{H_2(\bar F_Y(x))}\right]\text{ is increasing in }x>0,
\end{eqnarray*}
which holds if
\begin{eqnarray}\label{eq1}
\frac{r_X(x)}{r_Y(x)}\text{ is increasing in }x>0
\end{eqnarray}
and
\begin{eqnarray}\label{eq2}
\frac{H_1(\bar F_X(x))}{H_2(\bar F_Y(x))}\text{ is increasing in }x>0.
\end{eqnarray}
Note that \eqref{eq1} holds because $X\underset{ c}{\prec} Y$. Further, \eqref{eq2} holds if and only if
  \begin{eqnarray}\label{eq3}
\tilde r_Y(x)\left[(1-\bar F_Y(x))\frac{H_2'(\bar F_Y(x))}{H_2(\bar F_Y(x))} \right] \geq \tilde r_X(x)\left[(1-\bar F_X(x))\frac{H_1'(\bar F_X(x))}{H_1(\bar F_X(x))} \right],\text{ for all }x>0. 
  \end{eqnarray}
  Since $Y\leq_{rh} X $, we have
  \begin{eqnarray}\label{eq4}
  \tilde r_Y(x)\leq \tilde r_X(x) \text{ and }\bar F_Y(x)\leq \bar F_X(x),\text{ for all } x>0.
  \end{eqnarray}
  Now consider the following two cases.
  \\Case-I: Let $(1-p){H_1'(p)}/{H_1(p)}$ is decreasing in $p\in (0,1)$.
  Then
  \begin{eqnarray*}
  (1-\bar F_Y(x))\frac{H_2'(\bar F_Y(x))}{H_2(\bar F_Y(x))}&\geq & (1-\bar F_Y(x))\frac{H_1'(\bar F_Y(x))}{H_1(\bar F_Y(x))}\nonumber
  \\&\geq& (1-\bar F_X(x))\frac{H_1'(\bar F_X(x))}{H_1(\bar F_X(x))},\text{ for all } x>0,
  \end{eqnarray*}
  where the first inequality follows from condition ($i$), and the second inequality follows from \eqref{eq4} and condition~($ii$). 
   \\Case-II: Let $(1-p){H_2'(p)}/{H_2(p)}$ is decreasing in $p\in (0,1)$.
  Then
  \begin{eqnarray*}
  (1-\bar F_Y(x))\frac{H_2'(\bar F_Y(x))}{H_2(\bar F_Y(x))}&\geq & (1-\bar F_X(x))\frac{H_2'(\bar F_X(x))}{H_2(\bar F_X(x))}\nonumber
  \\&\geq& (1-\bar F_X(x))\frac{H_1'(\bar F_X(x))}{H_1(\bar F_X(x))},\text{ for all } x>0,
  \end{eqnarray*}
  where the first inequality follows from \eqref{eq4} and ($ii$), and the second inequality follows from~($i$). 
  Now, from Cases I and II, we get that
   \begin{eqnarray}\label{eq5}
  -(1-\bar F_Y(x))\frac{H_2'(\bar F_Y(x))}{H_2(\bar F_Y(x))}
  \leq -(1-\bar F_X(x))\frac{H_1'(\bar F_X(x))}{H_1(\bar F_X(x))},\text{ for all } x>0.
  \end{eqnarray}
  Further, ($i$) implies that
   \begin{eqnarray}\label{eq6}
   -(1-\bar F_X(x))\frac{H_1'(\bar F_X(x))}{H_1(\bar F_X(x))}\geq 0,\text{ for all } x>0.
  \end{eqnarray}
  On combing \eqref{eq4}, \eqref{eq5} and \eqref{eq6}, we get \eqref{eq3}. Hence, the result is proved.$\hfill\Box$
\\\\{\bf Proof of Theorem~\ref{t2}:} Note that
\begin{eqnarray*}
F_{\tau_1\left(\mbox{\boldmath$X$}\right)}(x)=1-h_1\left(\bar F_X(x)\right)\text{ and } F_{\tau_2\left(\mbox{\boldmath$Y$}\right)}(x)=1-h_2\left(\bar F_Y(x)\right),\quad x>0,
\end{eqnarray*}
which gives reversed failure rates of $\tau_1\left(\mbox{\boldmath$X$}\right)$ and $\tau_2\left(\mbox{\boldmath$Y$}\right)$ as 
\begin{eqnarray*}
\tilde r_{\tau_1\left(\mbox{\boldmath$X$}\right)}(x)=\frac{f_X(x)h_1'\left(\bar F_X(x)\right)}{1-h_1\left(\bar F_X(x)\right)}=\tilde r_X(x)R_1(\bar F_X(x)),\quad x>0
\end{eqnarray*}
and
\begin{eqnarray*}
\tilde r_{\tau_2\left(\mbox{\boldmath$Y$}\right)}(x)=\frac{f_Y(x)h_2'\left(\bar F_Y(x)\right)}{1-h_2\left(\bar F_X(x)\right)}=\tilde r_Y(x)R_2(\bar F_Y(x)),\quad x>0,
\end{eqnarray*}
 respectively. Then, $\tau_1\left(\mbox{\boldmath$X$}\right)\underset{ b}{\prec}\tau_2\left(\mbox{\boldmath$Y$}\right)$ holds if and only if
\begin{eqnarray*}
\frac{\tilde r_{\tau_1\left(\mbox{\boldmath$X$}\right)}(x)}{\tilde r_{\tau_2\left(\mbox{\boldmath$X$}\right)}(x)}=\left[\frac{\tilde r_X(x)}{\tilde r_Y(x)}\right]\left[\frac{R_1(\bar F_X(x))}{R_2(\bar F_Y(x))}\right]\text{ is decreasing in }x>0,
\end{eqnarray*}
which holds if
\begin{eqnarray}\label{eq21}
\frac{\tilde r_X(x)}{\tilde r_Y(x)}\text{ is decreasing in }x>0
\end{eqnarray}
and
\begin{eqnarray}\label{eq22}
\frac{R_1(\bar F_X(x))}{R_2(\bar F_Y(x))}\text{ is decreasing in }x>0.
\end{eqnarray}
Note that \eqref{eq21} holds because $X\underset{ b}{\prec} Y$. Further, \eqref{eq22} holds if and only if
  \begin{eqnarray}\label{eq23}
r_Y(x)\left[\bar F_Y(x)\frac{R_2'(\bar F_Y(x))}{R_2(\bar F_Y(x))} \right] \leq r_X(x)\left[\bar F_X(x)\frac{R_1'(\bar F_X(x))}{R_1(\bar F_X(x))} \right],\text{ for all } x>0. 
  \end{eqnarray}
  Since $X\leq_{hr} Y $, we have
  \begin{eqnarray}\label{eq24}
   r_Y(x)\leq r_X(x) \text{ and }\bar F_X(x)\leq \bar F_Y(x),\text{ for all } x>0.
  \end{eqnarray}
  Now consider the following two cases.
  \\Case-I: Let $p{R_1'(p)}/{R_1(p)}$ is decreasing in $p\in (0,1)$.
  Then
  \begin{eqnarray*}
  \bar F_Y(x)\frac{R_2'(\bar F_Y(x))}{R_2(\bar F_Y(x))}&\leq & \bar F_Y(x)\frac{R_1'(\bar F_Y(x))}{R_1(\bar F_Y(x))}\nonumber
  \\&\leq& \bar F_X(x)\frac{R_1'(\bar F_X(x))}{R_1(\bar F_X(x))},\text{ for all } x>0,
  \end{eqnarray*}
  where the first inequality follows from ($i$), and the second inequality follows from \eqref{eq24} and~($ii$). 
   \\Case-II: Let $p{R_2'(p)}/{R_2(p)}$ is decreasing in $p\in (0,1)$.
  Then
  \begin{eqnarray*}
  \bar F_Y(x)\frac{R_2'(\bar F_Y(x))}{R_2(\bar F_Y(x))}&\leq & \bar F_X(x)\frac{R_2'(\bar F_X(x))}{R_2(\bar F_X(x))}\nonumber
  \\&\leq& \bar F_X(x)\frac{R_1'(\bar F_X(x))}{R_1(\bar F_X(x))},\text{ for all } x>0,
  \end{eqnarray*}
  where the first inequality follows from \eqref{eq24} and ($ii$), and the second inequality follows from~($i$). 
  Now, from Cases I and II, we get that
   \begin{eqnarray}\label{eq25}
  \bar F_Y(x)\frac{R_2'(\bar F_Y(x))}{R_2(\bar F_Y(x))}
  \leq \bar F_X(x)\frac{R_1'(\bar F_X(x))}{R_1(\bar F_X(x))},\text{ for all } x>0.
  \end{eqnarray}
  Further, ($i$) implies that
   \begin{eqnarray}\label{eq26}
   \bar F_X(x)\frac{R_1'(\bar F_X(x))}{R_1(\bar F_X(x))}\geq 0,\text{ for all } x>0.
  \end{eqnarray}
  On combing \eqref{eq24}, \eqref{eq25} and \eqref{eq26}, we get \eqref{eq23}. Hence, the result is proved.$\hfill\Box$
\\\\{\bf Proof of Theorem~\ref{t41}:} We have
 \begin{eqnarray*}
 \bar F_{T_C}(x)=h\left(1-(1-\bar F_X(x))^{m+1}\right), \quad x>0
 \end{eqnarray*}
 and 
 \begin{eqnarray*}
  \bar F_{T_S}(x)=1-\left(1-h(\bar F_X(x))\right)^{m+1},\quad x>0,
 \end{eqnarray*}
 which gives failure rates of $T_C$ and $T_S$ as 
  \begin{eqnarray*}
  r_{T_C}(x)=(m+1)f_X(x)\left(1-\bar F_X(x)\right)^{m}\frac{h'\left(1-(1-\bar F_X(x))^{m+1}\right)}{h\left(1-(1-\bar F_X(x))^{m+1}\right)},\quad x>0
 \end{eqnarray*}
 and
 \begin{eqnarray*}
  r_{T_S}(x)=(m+1)f_X(x)\left(1-h(\bar F_X(x))\right)^{m}\frac{h'\left(\bar F_X(x)\right)}{1-\left(1-h(\bar F_X(x))\right)^{m+1}},\quad x>0,
 \end{eqnarray*}
  respectively. Then $T_S \underset{ c}{\prec}(\text{resp. }\underset{ c}{\succ})~ T_C$ holds if, and only if,
 \begin{eqnarray*}
\frac{ r_{T_S}(x)}{ r_{T_C}(x)} =\left(\frac{\left(1-h(\bar F_X(x))\right)^{m}h'\left(\bar F_X(x)\right)}{1-\left(1-h(\bar F_X(x))\right)^{m+1}}\right)\left(\frac{h\left(1-(1-\bar F_X(x))^{m+1}\right)}{\left(1-\bar F_X(x)\right)^{m}h'\left(1-(1-\bar F_X(x))^{m+1}\right)}\right)
 \end{eqnarray*}
is increasing (resp. decreasing) in $x>0$, which is equivalent to the fact that
 \begin{eqnarray*}
 \left(\frac{\left(1-h(p)\right)^{m}h'\left(p\right)}{1-\left(1-h(p)\right)^{m+1}}\right)\left(\frac{h\left(1-(1-p)^{m+1}\right)}{\left(1-p\right)^{m}h'\left(1-(1-p)^{m+1}\right)}\right)\nonumber
 \end{eqnarray*}
 is decreasing (resp. incresaing) in $p\in(0,1)$. Hence the result is proved.$\hfill\Box$
\\\\{\bf Proof of Corollary~\ref{co1}:} The reliability function of an $n$-out-of-$n$ system is given by
  $h(p)=p^n.$ 
  Thus, to prove the result it suffices to show that \eqref{km0} holds for $h(p)=p^n$ with $m=1$ and $n\geq 2$. Note that it holds if and only if
  \begin{eqnarray*}
  \frac{(2-p)(1-p^n)}{(1-p)(2-p^n)} \text{ is increasing in }p\in(0,1),
  \end{eqnarray*}
  or equivalently,
   \begin{eqnarray*}
  1+\zeta_1(p) \text{ is increasing in }p\in(0,1),
  \end{eqnarray*}
  where
  $$\zeta_1(p)=\frac{p-p^n}{2-2p-p^n+p^{n+1}},\quad 0<p<1.$$
  Now,
  \begin{eqnarray*}
  \zeta_1'(p)\stackrel{\text{sgn}}=2-\zeta_2(p),\quad 0<p<1,
  \end{eqnarray*}
  where
  $$\zeta_2(p)=2np^{n-1}-3(n-1)p^n+n p^{n+1}-p^{2n},\quad 0<p<1.$$
  This gives
  $$\zeta'_2(p)=np^{n-2}\zeta_3(p),\quad 0<p<1,$$
  where
  \begin{eqnarray*}
  \zeta_3(p)&=&2(n-1)-3(n-1)p+(n+1)p^{2}-2p^{n+1}
  \\&\geq &(n-1)(2-3p+p^2)
  \\&=&(n-1)(2-p)(1-p)\geq 0, \quad 0<p<1.
  \end{eqnarray*}
This implies that $\zeta_2(p)$ is increasing in $p\in (0,1)$ with $\zeta_2(0)=0$ and $\zeta_2(1)=2$, and hence $0\leq \zeta_2(p)\leq 2$. Again, this implies that $\zeta_1(p)$ is increasing in $p\in(0,1),$ and hence the result is proved.$\hfill\Box$
\\\\{\bf Proof of Theorem~\ref{t3}:} We have
 \begin{eqnarray*}
 F_{T_C}(x)=1-h\left(1-(1-\bar F_X(x))^{m+1}\right),\quad x>0
 \end{eqnarray*}
 and 
 \begin{eqnarray*}
  F_{T_S}(x)=\left(1-h(\bar F_X(x))\right)^{m+1},\quad x>0,
 \end{eqnarray*}
 which gives reversed failure rates of $T_C$ and $T_S$ as
  \begin{eqnarray*}
 \tilde r_{T_C}(x)=(m+1)f_X(x)\left(1-\bar F_X(x)\right)^{m}\frac{h'\left(1-(1-\bar F_X(x))^{m+1}\right)}{1-h\left(1-(1-\bar F_X(x))^{m+1}\right)},\quad x>0
 \end{eqnarray*}
 and
 \begin{eqnarray*}
 \tilde r_{T_S}(x)=(m+1)f_X(x)\frac{h'\left(\bar F_X(x)\right)}{1-h\left(\bar F_X(x)\right)},\quad x>0,
 \end{eqnarray*}
 respectively. Then $T_S \underset{ b}{\prec} T_C$ holds if and only if
 \begin{eqnarray*}
 \frac{\tilde r_{T_S}(x)}{\tilde r_{T_C}(x)}=\left(\frac{h'\left(\bar F_X(x)\right)}{1-h\left(\bar F_X(x)\right)}\right)\left(\frac{1-h\left(1-(1-\bar F_X(x))^{m+1}\right)}{\left(1-\bar F_X(x)\right)^{m}h'\left(1-(1-\bar F_X(x))^{m+1}\right)}\right)\text{ is decreasing in }x>0,
 \end{eqnarray*}
 or equivalently,
 \begin{eqnarray*}
 \left(\frac{(1-p)h'\left(p\right)}{1-h\left(p\right)}\right)\left(\frac{1-h\left(1-(1-p)^{m+1}\right)}{\left(1-\left(1-(1-p)^{m+1}\right)\right)h'\left(1-(1-p)^{m+1}\right)}\right)\text{ is increasing in }p\in(0,1).
 \end{eqnarray*}
 This is equivalent to the fact that
 $$\frac{R(p)}{R(1-(1-p)^{m+1})}\text{ is incresaing in }p\in(0,1),$$
 and hence the result is proved.$\hfill\Box$
\\\\{\bf Proof of Proposition~\ref{p41}:} Since $pR'(p)/R(p)$ is decreasing in $p\in(0,1)$, and $p\leq1-(1-p)^{m+1}$, for all $p\in(0,1)$, we have
  \begin{eqnarray}\label{cs2}
  p\frac{R'(p)}{R(p)}\geq \left(1-(1-p)^{m+1}\right)\frac{R'(1-(1-p)^{m+1})}{R(1-(1-p)^{m+1})
  }, \text{ for all }p\in(0,1).
  \end{eqnarray}
  Further, it can be easily checked that, for all $p\in(0,1),$ $$1-(1-p)^{m+1}\geq (m+1)p(1-p)^{m}.$$
  Since $pR'(p)/R(p)$ is positive for all $p\in(0,1)$, we get from the above inequality that,$\text{ for all }p\in(0,1)$,
   \begin{eqnarray}\label{cs3}
  \left(1-(1-p)^{m+1}\right)\frac{R'(1-(1-p)^{m+1})}{R(1-(1-p)^{m+1}}\geq (m+1)p(1-p)^{m}\frac{R'(1-(1-p)^{m+1})}{R(1-(1-p)^{m+1})}.
  \end{eqnarray}
On combining \eqref{cs2} and \eqref{cs3}, we get
\begin{eqnarray*}
  p\frac{R'(p)}{R(p)}\geq (m+1)p(1-p)^{m}\frac{R'(1-(1-p)^{m+1})}{R(1-(1-p)^{m+1}},\text{ for all }p\in(0,1),
  \end{eqnarray*}
  or equivalently, 
   \begin{eqnarray*}
 \frac{R(p)}{R(1-(1-p)^{m+1})}\text{ is increasing in }p\in(0,1),
 \end{eqnarray*}
 and hence the result follows from Theorem~\ref{t3}.$\hfill\Box$
\\\\{\bf Proof of Theorem~\ref{th1}:} Note that, for any fixed $t>0$,
\begin{eqnarray}\label{rs}
\bar F_{\tau\left(\mathbf{X}_{t}\right)}(x)=h\left(\frac{\bar F_X(t+x)}{\bar F_X(t)}\right)\text{ and }\bar F_{\left(\tau\left(\mathbf{X}\right)\right)_{t}}(x)=\frac{h\left(\bar F_X(t+x)\right)}{h\left(\bar F_X(t)\right)},\quad x>0,
\end{eqnarray}
which gives
\begin{eqnarray*}
r_{\tau\left(\mathbf{X}_{t}\right)}(x)=\frac{f_X(t+x)h'\left(\frac{\bar F_X(t+x)}{\bar F_X(t)}\right)}{\bar F_X(t)h\left(\frac{\bar F_X(t+x)}{\bar F_X(t)}\right)}=r_X(t+x)H\left(\frac{\bar F_X(t+x)}{\bar F_X(t)}\right),\quad x>0,
\end{eqnarray*}
and
\begin{eqnarray*}
r_{\left(\tau\left(\mathbf{X}\right)\right)_{t}}(x)=\frac{f_X(t+x)h'\left(\bar F_X(t+x)\right)}{h\left(\bar F_X(t+x)\right)}=r_X(t+x)H(\bar F_X(t+x)),\quad x>0.
\end{eqnarray*}
 Then, $\tau\left(\mathbf{X}_{t}\right)\underset{ c}{\prec}\left(\tau\left(\mathbf{X}\right)\right)_{t}$ holds if and only if
\begin{eqnarray*}
\frac{r_{\tau\left(\mathbf{X}_{t}\right)}(x)}{r_{\left(\tau\left(\mathbf{X}\right)\right)_{t}}(x)}=\frac{H\left(\frac{\bar F_X(t+x)}{\bar F_X(t)}\right)}{H(\bar F_X(t+x))}\text{ is increasing in }x>0,
\end{eqnarray*}
which is equivalent to the fact that, for all $q\in(0,1)$,
$$\frac{H\left(\frac{p}{q}\right)}{H(p)}\text{ is decreasing in }p\in(0,q).$$
Further, this holds if and only if
$$\frac{p}{q}\left(\frac{H'\left({p}/{q}\right)}{H(p/q)}\right)\leq p\frac{H'(p)}{H(p)}, \text{ for all }0<p\leq q<1,$$
which is equivalent to the fact that
$$p\frac{H'(p)}{H(p)}\text{ is decreasing in }p\in(0,1).$$
Hence, the result is proved.$\hfill\Box$
\\\\{\bf Proof of Theorem~\ref{th2}:} Let $t>0$ be fixed. From \eqref{rs}, we have
\begin{eqnarray*}
\tilde r_{\tau\left(\mathbf{X}_{t}\right)}(x)=\frac{f_X(t+x)h'\left(\frac{\bar F_X(t+x)}{\bar F_X(t)}\right)}{\bar F_X(t)\left(1-h\left(\frac{\bar F_X(t+x)}{\bar F_X(t)}\right)\right)},\quad x>0
\end{eqnarray*}
and
\begin{eqnarray*}
\tilde r_{\left(\tau\left(\mathbf{X}\right)\right)_{t}}(x)=\frac{f_X(t+x)h'\left(\bar F_X(t+x)\right)}{h\left(\bar F_X(t)\right)-h\left(\bar F_X(t+x)\right)},\quad x>0.
\end{eqnarray*}
 Then, $\tau_1\left(\mbox{\boldmath$X$}_t\right)\underset{ b}{\prec}(\text{resp. }\underset{ b}{\succ})\;\left(\tau_2\left(\mbox{\boldmath$X$}\right)\right)_t$ holds if and only if
\begin{eqnarray*}
\frac{\tilde r_{\tau\left(\mathbf{X}_{t}\right)}(x)}{\tilde r_{\left(\tau\left(\mathbf{X}\right)\right)_{t}}(x)}=\left[\frac{h'\left(\frac{\bar F_X(t+x)}{\bar F_X(t)}\right)}{\bar F_X(t)\left(1-h\left(\frac{\bar F_X(t+x)}{\bar F_X(t)}\right)\right)}\right]\left[\frac{h\left(\bar F_X(t)\right)-h\left(\bar F_X(t+x)\right)}{h'\left(\bar F_X(t+x)\right)}\right]
\end{eqnarray*}
is decreasing (resp. increasing) in $x>0$, which is equivalent to \eqref{rs0}. Hence, the result is proved.$\hfill\Box$
\\\\{\bf Proof of Corollary~\ref{c02}:} The reliability function of a $1$-out-of-$n$ system is given by
  $h(p)=1-(1-p)^n,\;0<p<1.$ 
  Thus, to prove the result it suffices to show that \eqref{rs0} holds for $h(p)=1-(1-p)^n,\;0<p<1.$ Note that this holds if and only if, for every fixed $q\in(0,1)$, 
  $$\frac{(1-p)^n-(1-q)^n}{(1-p)^{n-1}(q-p)}\text{ is increasing in }p\in(0,q),$$
  or equivalently,
   $$\zeta_5(y)\stackrel{\text{def.}}=\frac{y^n-1}{(y-1)y^{n-1}}\text{ is decreasing in }y>1.$$
  We have
   $$\zeta_5'(y)\stackrel{\text{sgn}}=y^{n-2}\zeta_6(y),\quad y>1,$$
   where
   $$\zeta_6(y)=-y^n+ny-(n-1),\quad y>1.$$
   Note that $\zeta_6(\cdot)$ is a decreasing function with $\zeta_6(1)=0$, and hence $\zeta_6(y)\leq 0$ for all $y>1$. Further, this implies that $\zeta_5(y)$ is decreasing in $y>1$. Hence the result is proved.$\hfill\Box$
\end{document}